\documentclass[epj]{svjour}
\usepackage{latexsym}
\usepackage{graphics}
\usepackage{mathrsfs,amsfonts,graphicx,cancel,color,bm,ulem}
\usepackage{amsmath}
\usepackage{amssymb}
\usepackage{revsymb}
\usepackage{hyperref}
\usepackage{cite}

\newcommand{\tr}{{\textrm{Tr}}}
\def \e{\textrm{e}}

\def \A{\mathcal{A}}

\def\ket#1{|{#1}\rangle}
\def\bra#1{\langle{#1}|}

\begin{document}
\title{Measuring quantumness: From theory to observability in interferometric setups}
\author{Leonardo Ferro\inst{1} \and Rosario Fazio\inst{2} \and Fabrizio Illuminati\inst{3} \and Giuseppe Marmo\inst{4} \and Saverio Pascazio\inst{5} \and Vlatko Vedral\inst{6}
}                     
\institute{INFN, Sezione di Napoli, I-80126 Napoli, Italy, and Morgan Stanley Analytics, H-1095, Budapest, Hungary \and 
Abdus Salam ICTP, Strada Costiera 11, I-34151 Trieste, Italy, and NEST, Scuola Normale Superiore \& Istituto Nanoscienze-CNR, I-56126 Pisa, Italy
\and Dipartimento di Ingegneria Industriale, Universit\`a di Salerno, I-84084 Fisciano (SA), Italy, and INFN, Sezione di Napoli, I-80126 Napoli, Italy \and 
Dipartimento di Fisica, Universit\`a di Napoli, I-80126 Napoli, Italy, and INFN, Sezione di Napoli, I-80126 Napoli, Italy  \and Dipartimento di Fisica and MECENAS, Universit\`a di Bari, I-70126 Bari, Italy, Istituto Nazionale di Ottica (INO-CNR), I-50125 Firenze, Italy, and INFN, Sezione di Bari, I-70126 Bari, Italy
\and Clarendon Laboratory, Department of Physics, University of Oxford, Parks Road, Oxford OX1 3PU, United Kingdom, Centre for Quantum Technologies, National University of Singapore, 3 Science Drive 2, 117543, Singapore, and Department of Physics, National University of Singapore, 2 Science Drive 3, 117542, Singapore  
}
\date{Received: date / Revised version: date}
%
\abstract{
We investigate the notion of quantumness based on the non-commutativity of the algebra of observables and introduce a measure of quantumness based on the mutual incompatibility of quantum states. We show that such a quantity can be experimentally measured with an interferometric setup and that, when an arbitrary bipartition of a given composite system is introduced, it detects the one-way quantum correlations restricted to one of the two subsystems. We finally show that, by combining only two projective measurements and carrying out the interference procedure, our measure becomes an efficient universal witness of quantum discord and non-classical correlations.
\PACS{
      {03.65.-w}{Quantum mechanics}   \and
      {03.67.Mn}{Entanglement measures, witnesses, and other characterizations}
     } 
} 
\maketitle
\section{Introduction}
\label{intro}
A fundamental question in modern physics is how to characterize the crossover between the quantum and the classical world. There are many different approaches to the resolution of this problem, yet so far none can claim to capture all the many complex aspects of the question. Recently, much interest has been devoted to the exploration of the properties of composite quantum systems exhibiting non-classical correlations, with applications to a variety of fields \cite{review-entanglement1,review-entanglement2}. In this context, classical states are identified as those states whose correlations can be described in terms of classical probabilities. In light of this, different measures have been proposed to quantify non-classical correlations such as entanglement, discord, and related measures \cite{review-discord}.

An alternative approach to characterize the crossover, based on the definition of the quantumness of a single physical system, focuses on the non-commutativity of the algebra of observables \cite{alicki2008sem,alicki2008quantumness,FPVY12,FFMP14}. In this framework a system is defined to be classical if all its  accessible states commute with each other \cite{fazio2013witnessing}. The advantage of this intrinsic approach is that it does not depend on an arbitrary choice of bipartition of the system. This aspect becomes particularly relevant, for instance, in the case of identical particles, for which one cannot rely on the tensor product structure of Hilbert spaces in order to quantify entanglement and quantum correlations \cite{marmo2011,benatti2012,balachandran2013}. 
Recently, this approach has found applications in quantum technologies \cite{pfeiffer,sanz}, in 
the analysis of quantum coherence and correlations \cite{hufan,bhattacharya}, and in the characterization of the fundamental quantum speed limits governing the generation of nonclassicality and the mutual incompatibility of two states connected by an arbitrary physical process \cite{jing,mirkin,liu}. 
Moreover, the intrinsic algebraic approach paves the way to applications to different physical situations in which it is unfeasible to compute correlations between parties that are in principle distinguishable, but actually extremely hard to discriminate in practice, such as in complex many-body interacting systems and in biological matter \cite{biology}.

In this Article we will address the problem of characterizing the quantum nature of a system in terms of the degree of non-commutativity of quantum states, in an operational way suitable to experimental verification. As shown in \cite{fazio2013witnessing}, the non-commutativity of two states can be witnessed by relying on the anti-commutator of the states, for which an experimental verification scheme is in principle available. However, the experimental procedure turns out to depend on a state-dependent iterative procedure. For some states, a large number of copies is required in order to characterize precisely their quantum properties.

In the following we propose a witness for the global quantum properties of a state, and provide a \textit{universal} quantum circuit for the experimental verification of such characterization, that is independent of the input states. The main result will be that it is always possible to determine the global quantum properties of a state by setting up a quantum interference experiment involving only two copies of the input states.

Furthermore, as an interesting spin-off, we will show that our measure of quantumness and incompatibility can also be used to experimentally detect quantum correlations between two arbitrary parties in partition-dependent settings, thus providing a general method to witness quantum discord and related measures of quantum correlations.

\section{Witnessing and quantifying quantumness}
\label{sec:wit}
Given two states $\rho_a$ and $\rho_b$, we can quantify their mutual incompatibility as twice the Hilbert-Schmidt norm of their commutator:
\begin{eqnarray}
\label{quantumness_measure}
 Q(\rho_a,\rho_b) = 2 \|[\rho_a,\rho_b]\|^2 &=& 2 \tr([\rho_a,\rho_b]^2)\nonumber \\
 &=& 4 \tr\left((\rho_a\rho_b)^2 -\rho_a^2\rho_b^2 \right).
\end{eqnarray}
The key observation is that the trace of a positive operator is positive and vanishes if and only if the operator is null. For this reason, $Q$ turns out to be a very powerful \textit{quantumness witness}, since $Q(\rho_a,\rho_b) = 0$ if and only if $[\rho_a,\rho_b] = 0$. In particular it is straightforward to verify that  \cite{iyengar2013quantifying}
\begin{equation}
 0 \leq Q(\rho_a,\rho_b) \leq 1.
\end{equation}
In order to motivate this definition and put it in a proper context, it is useful to add a few remarks.
Let $\A$ be the algebra of the observables of the system \cite{von-Neumann-book}. We say that a state $\rho$ is \textit{classical} if and only if \cite{FPVY12,FFMP14}
\begin{equation}
\tr( \rho\, [A, B] ) = 0, \qquad \forall\, A,B \in \A.
 \label{classst}
\end{equation}
In words, we say that a state is classical if it does not detect the presence of non-vanishing commutators in the full algebra of observables. Otherwise, the state is \textit{quantum}.
We emphasize that the above definition of classical states depends on the algebra of observables (namely, from the  physical point of view, on which observables are experimentally accessible). The same state can be classical or quantum depending on $\A$ \cite{FPVY12,FFMP14}. A few examples will help elucidating this point. 

\textit{Example 1.}
Consider a qubit and let $\A$ be the algebra of its observables. According to definition
(\ref{classst}), a state is said to be classical if it does not detect the presence of non-vanishing commutators in $\A$. Otherwise, the state is quantum.
Consider an experimental situation in which one has only \textit{experimental access} to the algebra $\A_1$ generated by $\{\openone, \sigma_z\}$. Then the state $\rho=p\ket{0}\bra{0}+q\ket{1}\bra{1}$ is classical. Consider now an experimental situation in which one has \textit{experimental access} to the full algebra $\A_2=\mathfrak{u}(2)$, generated by $\{\openone, \sigma_x, \sigma_y, \sigma_z\}$. Then the state $\rho=p\ket{0}\bra{0}+q\ket{1}\bra{1}$ is quantum. Indeed, one can bring to light coherence, e.g.\ $\bra{-}\rho\ket{+}=c_0c_1(p-q)$ for $\ket{+}=c_0\ket{0}+c_1\ket{1}$ and $\ket{-}=c_1^*\ket{0}-c_0^*\ket{1}$,  which is nonvanishing provided $p\neq q$ and $c_0,c_1\neq0$. This is what one means when one says that, in general, mixtures like $\rho$ are not classical states. 

On the other hand, the completely mixed state $\rho=\openone/2$ is classical for any algebra $\A$, in that it does not possess any coherence, $\bra{-}\rho\ket{+}=0$ for any $c_0$ and $c_1$.
Therefore, the definition of classicality turns out to be heavily dependent on the algebra of observables, namely on the experimental situation. If the experiment does not allow measurements of, say, $\sigma_x$ and $\sigma_y$, a state like 
$\rho=p\ket{0}\bra{0}+q\ket{1}\bra{1}$ is classical. If, by contrast, measurements of both $\sigma_x$ and $\sigma_y$ are feasible, the same state is quantum.

\textit{Example 2.}
Let us look at another example, that is more involved and will turn out to be useful in the following.
Consider a composite system in a quantum-classical state \cite{OZ,HV} 
\begin{equation}
\label{quantum_classical1}
\rho_{AB}^{QC} = \displaystyle\sum_{i=1}^n p_i \rho_i \otimes \ket{b_i}\bra{b_i}.
\end{equation}
This state exhibits only classical correlations. Nevertheless with respect to the whole algebra of observables $\mathfrak{u}(n) \otimes \mathfrak{u}(n)$ it is a quantum state as it clearly displays quantum features (quantum coherence is a basis-dependent concept).
However, if we consider a different experimental situation in which the available algebra of observables  $\A$ is generated by 
\begin{equation}
	\{ \openone \otimes \ket{b_i}\bra{b_i},\ i=1,\ldots,n\},
\end{equation}
then the state $\rho_{AB}^{QC}$ cannot bring to light any quantum property and is  therefore classical. 
Thus, our definition (\ref{classst}) of classical (and quantum) states goes beyond the focus on correlations, and depends on the experimental situation, namely the algebra of observables. In the following sections we shall indeed focus on the experimental detection of quantumness via an interferometric setup.

Before we proceed towards experimental setups, let us add a few more comments.
An intuitive understanding would suggest that if two states $\rho_a$ and $\rho_b$ do not commute, then they are quantum; if two states $\rho_a$ and $\rho_b$ are classical then they commute. We now rigorously clarify the connection between the incompatibility of two states and their global quantum properties, referred to the whole algebra of observables. The first observation is the following:
\begin{equation}
	[\rho_a,\rho_b] \neq 0 \quad \Longrightarrow \quad \rho_a\ \textrm{and}\ \rho_b\ \textrm{are quantum states}.
\label{th1}
\end{equation}
Indeed, according to the definition (\ref{classst}), in order to prove that a state $\rho$ is quantum, we need to show that there exist observables $A, B$ such that $\tr (\rho\, [A,B]) \neq 0$. Note that to each state we can associate an observable in $\mathcal{A}$ having the density matrix as a matrix representation. Then by setting $A = i\,[\rho_a,\rho_b]$ and $B = \rho_b$, we have
\begin{equation}\label{witness1}
	\tr(\rho_a\, [A,B]) = -i \displaystyle \frac{Q(\rho_a,\rho_b)}{2} \neq 0,
\end{equation}
Alternatively, by setting $B = \rho_a$ we get
\begin{equation}\label{witness2}
	\tr(\rho_b\, [A,B]) = \displaystyle i \frac{Q(\rho_a,\rho_b)}{2} \neq 0,
\end{equation}
which proves (\ref{th1}). Equations (\ref{witness1}) and (\ref{witness2}) imply that $Q$ is a proper witness of the quantum nature of the states $\rho_a$ and $\rho_b$.
An assertion equivalent to~(\ref{th1}) is that if $\rho_a$ is classical then $[\rho_a,\rho_b] = 0\ \forall\, \rho_b$.

These remarks show that our measure $Q$ not only quantifies the relative quantumness of two given states, but is also a witness of the global quantum nature of the states, as $Q(\rho_a,\rho_b) > 0$ implies that both $\rho_a$ and $\rho_b$ are quantum states, by virtue of Proposition (\ref{th1}). At the same time, owing to general theorems \cite{FPVY12,FFMP14}, it yields information about the quantum structure of the observables of the system.

\section{Measuring and detecting quantumness}
\label{sec:meas}
\begin{figure}
\resizebox{0.49\textwidth}{!}{%
  \includegraphics{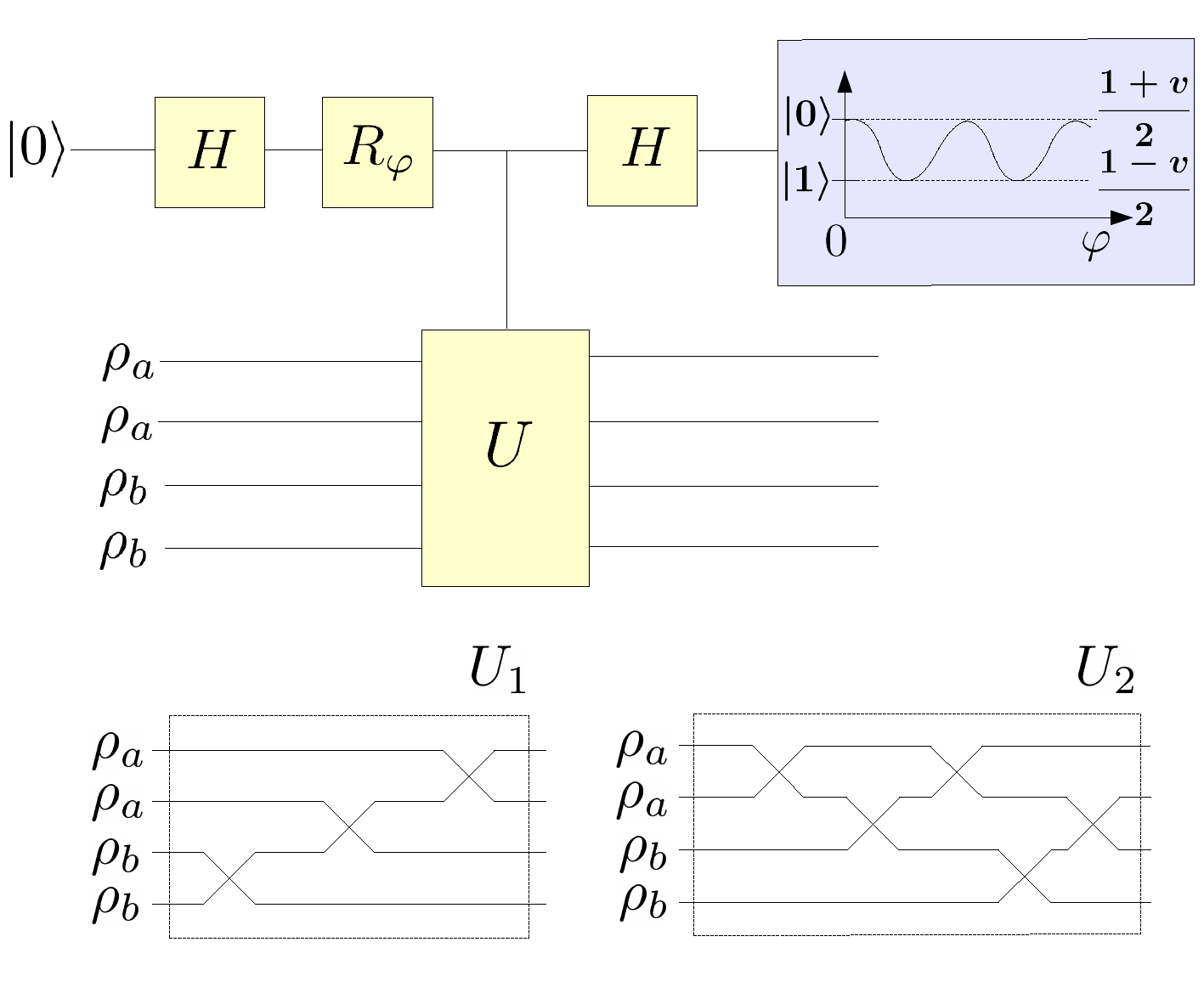}
}
\caption{(Color online) The relative quantumness of states $\rho_a$ and $\rho_b$ is detected by two quantum circuits of the type shown in the upper panel,
through the visibilities of the interference patterns of the control qubit, according to Eq.\ (\ref{quantumness_measure}). $H$ is the
Hadamard gate, $R_\varphi$ the phase-shift gate and $U$ the (controlled-)$U$ gate.
Lower panel, left: the unitary gate $U_1$ yields visibility $v_1 =  \tr(\rho_a^2\rho_b^2)$. Lower panel, right: the unitary gate $U_2$ yields visibility
$v_2 =  \tr\left((\rho_a\rho_b)^2\right)$. The relative quantumness is the difference of the two visibilities $Q(\rho_a,\rho_b) = 4 (v_2-v_1)$.
}
\label{fig:controlled-U}       
\end{figure}

The quantity $Q(\rho_a,\rho_b)$ in Eq.\ (\ref{quantumness_measure}) can be measured by using the interferometric setup displayed in Fig.\ \ref{fig:controlled-U}. The input state is $\rho_{\textrm{in}} = \ket 0\bra 0 \otimes \rho = \ket 0\bra 0 \otimes \rho_a \otimes \rho_a \otimes \rho_b \otimes \rho_b$, where the qubit $\ket 0$ controls the unitary operation $U$ on the state $\rho$. In general, the action of the control gate $U$ modifies the interference pattern of the control qubit by the factor \cite{sjoqvist2000geometric,filip,Carteret,EAOHHK}
\begin{equation}
 \tr(U \rho) = v \e^{i\alpha},
\label{visib}
\end{equation}
where the visibility $v$ and phase shift $\alpha$ of the interference fringes depend on $U$. The observed modification of the visibility yields an estimate of $\tr(U \rho)$, i.e.\ the expectation value of the unitary operator $U$ on the state $\rho$.
The evaluation of $Q$ in (\ref{quantumness_measure}) requires the use of the above-mentioned scheme for the measurement of the quantities
\begin{equation}
 \tr\left((\rho_a\rho_b)^2\right) \quad \textrm{and} \quad \tr(\rho_a^2\rho_b^2)
\end{equation}
in two separate experiments, or in the same interferometer by using two control qubits. Such operations can be easily constructed by cascading different
\textsc{swap} operators. We define the (unitary) generalized \textsc{swap} operator $S_{ij}$
\begin{eqnarray}
 & & S_{ij} \left\{ \ket{\phi_1} \otimes \ket{\phi_2} \cdots \otimes \ket{\phi_i} \cdots \otimes \ket{\phi_j} \cdots \right\} \nonumber \\
 & & \qquad = \ket{\phi_1} \otimes \ket{\phi_2} \cdots \otimes \ket{\phi_j} \cdots \otimes\ket{\phi_i} \cdots
\end{eqnarray} 
that exchanges ket $i$ with ket $j$. If we denote by $A,B,C,D$ the four parties of the state $\rho = \rho_a \otimes \rho_a \otimes \rho_b \otimes \rho_b$, then we have
\begin{equation}\label{v1}
v_1 = \tr(\rho_a^2\rho_b^2) = \tr(S_{AB}S_{BC}S_{CD}\ \rho_a \otimes \rho_a \otimes \rho_b \otimes \rho_b)
\end{equation}
and
\begin{eqnarray}
v_2 &= & 
\tr\left((\rho_a\rho_b)^2\right) \label{v2}  \\
 & = & \tr(S_{BC}S_{CD}S_{AB}S_{BC}S_{AB}\ \rho_a \otimes \rho_a \otimes \rho_b \otimes \rho_b) \qquad \nonumber
\end{eqnarray}
(there are in fact several equivalent circuits yielding the same effect).
See the lowers panels in Fig.\ \ref{fig:controlled-U}.
In both cases the quantity in Eq.\ (\ref{visib}) is real and Eq.\ (\ref{quantumness_measure}) yields
\begin{equation}
\label{Qvis}
 Q(\rho_a,\rho_b) = 4 (v_2-v_1).
\end{equation}
This is the result we sought: if $Q >0$, states $\rho_1$ and $\rho_2$ do not commute and the system cannot be classical.

It is worth remarking here that this interferometric setup realizes a significant generalization of the Hong-Ou-Mandel interferometric scheme \cite{Hong-Ou-Mandel1987}, in that our visibilities (\ref{v1}),\,(\ref{v2}) correspond to non-linear, bi-quadratic functions of the states that go beyond the bi-linear functions that enter in the Hong-Ou-Mandel visibility $V = \tr ( \rho_a \rho_b )$.

\section{Detection of quantum correlations}
\label{sec:detect}
Besides its intrinsic, partition-independent meaning, the global mutual incompatibility $Q$ bears important consequences for the detection of quantum correlations when bipartitions are introduced. Indeed, many efforts have been devoted to introduce experimental schemes for the detection of bipartite quantum correlations \cite{Experiment-scheme}, and optimal strategies have been found for the case of $2 \times d$ systems \cite{Girolami-Adesso-2012}. One-way quantum correlations have been observed in several experiments focused on particular quantum systems, such as a qubit encoded in a trapped ion in interaction with an environment \cite{Experiment-Nature-2014},  four qubits in a nuclear magnetic resonance processor \cite{Experiment-NMR-2011},  two-mode squeezed thermal states \cite{Experiment-Squeezed-2012} and pairs of polarization qubits \cite{Experiment-polarization-2013}. We will show now that by exploiting the properties of the measure $Q$ we are able to detect quantum correlations in arbitrary states of bipartite quantum systems $AB$.
A bipartite state $\rho_{AB}$ is said to exhibit classical correlations \cite{OZ,HV} if there exists some orthogonal basis $\ket{b_i}$ for party $B$ such that it acquires the so called quantum-classical form (\ref{quantum_classical1}), that we re-write here for convenience
\begin{equation}
\label{quantum_classical}
\rho_{AB}^{QC} = \displaystyle\sum_i p_i \rho_i \otimes \ket{b_i}\bra{b_i}.
\end{equation}
All the states that are not of this form are said to be quantum correlated.

Suppose Alice and Bob share a state and they want to check if they are quantum correlated. It is clear that any POVM measurement $\{E^i_A\}$ Alice can perform will leave the classical state (\ref{quantum_classical}) in a diagonal form in Bob's subsystem. Namely
\begin{equation}\label{conditional}
	\rho_{B|i} = \frac{\tr_A[(E^i_A \otimes \mathbb{I}_B) \rho_{AB}]}{\tr[(E^i_A \otimes \mathbb{I}_B) \rho_{AB}]}
\end{equation}
is a diagonal state, in the same basis of $B$, for any $E^i_A$. This implies that if Alice and Bob are classically correlated, all the conditional states of Bob will have a common eigenbasis, i.e.\ $[\rho_{B|i},\rho_{B|j}] = 0,\ \forall\, E^i_A,\, E^j_A$  \cite{chen2011detecting}.
Conversely if the conditional states of Bob will all commute for any local measurement of Alice, then the shared state $\rho_{AB}$ will be of the classical form (\ref{quantum_classical}). In conclusion,
\begin{equation}	
\rho_{AB} = \rho_{AB}^{QC} \ \Longleftrightarrow \ Q(\rho_{B|i}, \rho_{B|j})=0, \ \forall\,  E^i_A,\, E^j_A.
\end{equation}
If Alice and Bob are quantum correlated, then Alice can make measurements on her system such that the corresponding conditional states of Bob do not commute. Once Alice remotely prepares two noncommuting states $\rho_{B|1}$ and $\rho_{B|2}$ for Bob, then she simply has to communicate via a classical channel with him, who can measure the commutator of these states. In other words, Bob can carry out the procedure outlined in Fig.\ \ref{fig:controlled-U}. The measure $Q(\rho_{B|1}, \rho_{B|2})$ will be strictly positive if and only if Bob was quantum correlated to Alice:
\begin{equation}	
\rho_{AB} \neq \rho_{AB}^{QC} \Longleftrightarrow \exists\, E^1_A,\, E^2_A\ \textrm{s.t.}\ Q(\rho_{B|1}, \rho_{B|2}) > 0.
\end{equation}
This means that $Q$ turns out to be also a powerful quantum discord witness, since it suffices to perform two  measurements and carry out the interference procedure of Fig.\ \ref{fig:controlled-U} to reveal the quantum discord of any arbitrary bipartite quantum state.
Let us illustrate the above procedure with two paradigmatic examples.

\textit{Example 3.}
Consider the maximally-entangled EPR state
\begin{equation}
	{\ket{\psi_{AB}}}=\displaystyle\frac{1}{\sqrt{2}}\left({\ket 0}_A\otimes{\ket 0}_B + {\ket 1}_A\otimes{\ket 1}_B\right),
\end{equation}
and the two local measurements performed by $A$
\begin{equation}\label{operationsA}
	E^1_A = \Pi^1_A(\theta) \otimes \mathbb{I}_B,\quad E^2_A = \Pi^2_A(\theta,\varphi)\otimes\mathbb{I}_B,
\end{equation}
where $\Pi^1_A(\theta) = \ket{\psi_1(\theta)}\bra{\psi_1(\theta)}$, with
\begin{equation}
	\ket{\psi_1(\theta)} = \cos(\theta) \ket 0 + \sin(\theta) \ket 1,
\end{equation}
and $\Pi^2_A(\theta,\varphi) = \ket{\psi_2(\theta,\varphi)}\bra{\psi_2(\theta,\varphi)}$, with
\begin{equation}
	\ket{\psi_2(\theta,\varphi)} = \cos(\varphi)\ket{\psi_1(\theta)} + \sin(\varphi)\ket{\psi_1^\bot(\theta)},
\end{equation}
and
\begin{equation}
	\ket{\psi_1^\bot(\theta)} = \sin(\theta)\ket 0 - \cos(\theta) \ket 1.
\end{equation}
The conditional states of $B$ (\ref{conditional}) $\rho_{B|1}$ and $\rho_{B|2}$
do not commute in general, and our experimentally-detectable measure of quantumness (\ref{quantumness_measure}) gives $Q(\rho_{B|1},\rho_{B|2}) = \left(\sin{2\phi}\right)^2$, which is maximal at $\phi = \pi/4$, yielding $Q_{\textrm{max}} = 1$.

\textit{Example 4.}
Consider now the separable state
\begin{eqnarray}
\sigma_{AB} = \displaystyle &\frac{1}{4}&[ \ket 0\bra 0 \otimes \ket + \bra + + \ket 1 \bra 1 \otimes \ket - \bra - + \nonumber \\ &+& \ket + \bra + \otimes \ket 1 \bra 1 + \ket - \bra - \otimes \ket 0 \bra 0 ].
\end{eqnarray}
Obviously $\sigma_{AB}$ is not of the form (\ref{quantum_classical}) and therefore, although separable, it exhibits quantum correlation.
Let the party $A$ perform again the two local measurements (\ref{operationsA}) $E^1_A$ and $E^2_A$ and the conditional states of $B$ be $\sigma_{B|1}$ and $\sigma_{B|2}$. Then our quantumness witness (\ref{quantumness_measure}) detects the quantum correlations between $A$ and $B$ as $Q(\sigma_{B|1},\sigma_{B|2}) = (\sin{2\phi})^2/16$, which is again maximized by $\phi = \frac{\pi}{4}$, yielding $Q_{\textrm{max}} = \frac{1}{16}$.

\section{Conclusions and comments}
\label{sec:concl}
Let us briefly summarize and comment on the results obtained in this Article. We have introduced the global mutual incompatibility as a measure of the intrinsic quantum nature of physical states. Furthermore, we have introduced an interferometric method that detects the quantumness of a given system: given two states $\rho_a$ and $\rho_b$, the interferometer in Fig.\ \ref{fig:controlled-U} is able to check whether $[\rho_a,\rho_b] \neq 0$, detecting and quantifying their relative quantumness $Q$ (\ref{quantumness_measure}). The proposed scheme is universal and does not depend on the input states.

The method has application in the detection of quantum correlations. Consider two quantum systems that are correlated as in Eq.\ (\ref{quantum_classical}). Let one of the two parties (Alice) perform \textit{any} POVM. Then the states of the other party (Bob) will always commute. Conversely, if Alice can perform a POVM such that the commutator of Bob's states does not vanish, then the correlations have a quantum origin. 
It is sufficient for Alice to perform only two projective measurements in order to detect the existence of quantum correlations with Bob by letting him measure the quantumness witness $Q$ on the conditional states of his subsystem. The search for the two optimal projections is an interesting problem which will be investigated separately, but since the setup in Fig.\ \ref{fig:controlled-U} makes use of a rather simple interferometric method and can be applied to arbitrary $n \times m$ dimensional states, the scheme is in principle less demanding than other procedures explored in the literature \cite{Experiment-scheme,Girolami-Adesso-2012,Experiment-Nature-2014,Experiment-NMR-2011,Experiment-Squeezed-2012,Experiment-polarization-2013}.

In conclusion, we have discussed the concept of classical and quantum states, based on a measure of their incompatibility. Moreover, independently on possible interpretations, we have pursued a twofold objective: we have shown that the commutator of two states can be \textit{experimentally measured} via the interferometric setup of Fig.\ \ref{fig:controlled-U}, and moreover, such commutator is able to detect the quantum correlations of arbitrary $n \times m$ dimensional bipartite states. This result is a very general one and is  also easily experimentally accessible.

%

%
%
\section{Authors contributions}
All the authors were involved in the preparation of the manuscript.
All the authors have read and approved the final manuscript.
SP is partly supported by INFN through the project ``QUANTUM".

\end{document}